\documentclass[conference]{IEEEtran}
\IEEEoverridecommandlockouts
\usepackage{cite}
\usepackage{amsmath,amssymb,amsfonts}
\usepackage{stmaryrd}
\usepackage{mathrsfs}
\usepackage{algorithmic}
\usepackage{graphicx}
\usepackage{textcomp}
\usepackage{xcolor}
\usepackage[braket,qm]{qcircuit}
\usepackage{hyperref}
\usepackage{flushend}
\usepackage{fancyhdr}

\def\BibTeX{{\rm B\kern-.05em{\sc i\kern-.025em b}\kern-.08em
    T\kern-.1667em\lower.7ex\hbox{E}\kern-.125emX}}

\newcommand{\cn}{\ensuremath{\mathbb{C}}}
\newcommand{\rn}{\ensuremath{\mathbb{R}}}
\newcommand{\zn}{\ensuremath{\mathbb{Z}}}
\newcommand{\nn}{\ensuremath{\mathbb{N}}}

\newcommand{\Pp}{\ensuremath{\mathcal{P}}}
\newcommand{\Sp}{\ensuremath{\mathcal{S}}}
\newcommand{\Mp}{\ensuremath{\mathcal{M}}}
\newcommand{\Dp}{\ensuremath{\mathcal{D}}}

\newcommand{\ex}{\exists}

\renewcommand{\d}{\ensuremath{\partial}}
\renewcommand{\l}{\ensuremath{\left}}
\renewcommand{\r}{\ensuremath{\right}}
\newcommand{\lnorm}{\left\lVert}
\newcommand{\rnorm}{\right\lVert}

\newcommand{\norm}[1]{\lnorm {#1}\rnorm}

\newcommand{\rank}{\ensuremath{\mathrm{rank}}}
\newcommand{\init}{\ensuremath{\mathrm{init}}}
\newcommand{\opt}{\ensuremath{\mathrm{opt}}}
\newcommand{\lin}{\ensuremath{\mathrm{lin}}}
\newcommand{\vol}{\ensuremath{\mathrm{vol}}}
\newcommand{\Cost}{\ensuremath{\mathrm{Cost}}}
\renewcommand{\mod}{\ensuremath{\ \mathrm{mod}\ }}
\newcommand{\order}{\ensuremath{\mathcal{O}}}

\renewcommand{\phi}{\varphi}

\renewcommand{\rho}{\varrho}

\renewcommand{\theta}{\vartheta}
\newcommand{\eps}{\varepsilon}
\newcommand{\approxpropto}{\mathrel{\vcenter{\offinterlineskip\halign{\hfil$##$\cr\propto\cr\noalign{\kern2pt}\sim\cr\noalign{\kern-2pt}}}}}

\begin{document}

\title{Best-approximation error for \mbox{parametric quantum circuits} 
  \thanks{Research at Perimeter Institute is supported in part by the Government of Canada through the Department of Innovation, Science and Industry Canada and by the Province of Ontario through the Ministry of Colleges and Universities. S.K. acknowledges financial support from the Cyprus Research and Innovation Foundation under project ”Future-proofing Scientific Applications for the Supercomputers of Tomorrow (FAST)”, contract no. COMPLEMENTARY/0916/0048. M.S. and P.S. thank the Helmholtz Einstein International Berlin Research School in Data Science (HEIBRiDS) for funding.}
}

\author{
\IEEEauthorblockN{Lena Funcke}
\IEEEauthorblockA{\textit{Perimeter Institute for Theoretical Physics}\\
Waterloo, ON, Canada \\
lfuncke@perimeterinstitute.ca}
\and
\IEEEauthorblockN{Tobias Hartung}
\IEEEauthorblockA{\textit{Department of Mathematical Sciences} \\
\textit{University of Bath}\\
Bath, United Kingdom \\
th2040@bath.ac.uk}
\and
\IEEEauthorblockN{Karl Jansen}
\IEEEauthorblockA{\textit{NIC, DESY Zeuthen}\\
Zeuthen, Germany \\
Karl.Jansen@desy.de}
\and
\IEEEauthorblockN{Stefan Kühn}
\IEEEauthorblockA{\textit{Computation-Based Science and Technology Research Center} \\
\textit{The Cyprus Institute}\\
Nicosia, Cyprus \\
s.kuehn@cyi.ac.cy}
\and
\IEEEauthorblockN{Manuel Schneider}
\IEEEauthorblockA{\textit{NIC, DESY Zeuthen}\\
Zeuthen, Germany \\
manuel.schneider@desy.de}
\and
\IEEEauthorblockN{Paolo Stornati}
\IEEEauthorblockA{\textit{NIC, DESY Zeuthen}\\
Zeuthen, Germany \\
paolo.stornati@desy.de}
}

\lhead{To appear in the Proceedings of the 2021 IEEE International Conference on Web Services (ICWS), \copyright 2021 IEEE}
\maketitle
\thispagestyle{fancy}

\begin{abstract}
  In Variational Quantum Simulations, the construction of a suitable parametric quantum circuit is subject to two counteracting effects. The number of parameters should be small for the device noise to be manageable, but also large enough for the circuit to be able to represent the solution. Dimensional expressivity analysis can optimize a candidate circuit considering both aspects. In this article, we will first discuss an inductive construction for such candidate circuits. Furthermore, it is sometimes necessary to choose a circuit with fewer parameters than necessary to represent all relevant states. To characterize such circuits, we estimate the best-approximation error using Voronoi diagrams. Moreover, we discuss a hybrid quantum-classical algorithm to estimate the worst-case best-approximation error, its complexity, and its scaling in state space dimensionality. This allows us to identify some obstacles for variational quantum simulations with local optimizers and underparametrized circuits, and we discuss possible remedies.
\end{abstract}


\section{Introduction}
Noisy intermediate-scale quantum (NISQ) computers~\cite{Preskill2018} are opening up a new avenue to address a large class of computational problems that cannot be solved efficiently with classical computers. Applications of quantum computing range from machine learning~\cite{Biamonte2017} to finance~\cite{Orus2019} to various optimization problems~\cite{Montanaro2016,Brandao2017}. In physics, quantum computers intrinsically circumvent the sign problem that prevents Monte Carlo simulations of strongly correlated quantum-many body problems in certain parameter regimes~\cite{Troyer2005}. Although current hardware is of limited size and suffers from a considerable level of noise, the ability for NISQ devices to outperform classical computers has already been demonstrated successfully~\cite{Arute2019,Zhong2020} and techniques for mitigating the effects of noise are rapidly developing~\cite{Temme2017,Kandala2017,Endo2018,YeterAydeniz2020,Funcke2020}.

Many algorithms designed for NISQ devices make use of parametric quantum circuits. Variational quantum simulations~(VQSs)~\cite{Peruzzo2014,McClean2016}, a class of hybrid quantum-classical algorithms for solving optimization problems, are a particularly important example. Using a parametric quantum circuit, i.e., a quantum circuit composed of parameter dependent gates, a cost function is evaluated efficiently for a given set of parameters using the quantum coprocessor. The cost function is then minimized  on a classical computer in a feedback loop based on the measurement outcome obtained from the quantum device. Using cost functions related to the energy of the quantum state prepared on the quantum device, VQSs have been successfully applied to quantum many-body systems in quantum chemistry~\cite{Peruzzo2014,Wang2015,Kandala2017,Hempel2018} and even quantum mechanics and quantum field theory~\cite{Kokail2018,Hartung2018,Jansen2019,Paulson2020,Haase2020}.

Since VQSs depend on the choice of parametric quantum circuits, there are many open questions related to finding a good or optimal quantum circuit. For example, in order to be able to find the solution -- or at least find a good approximation -- a parametric quantum circuit needs to have many parameters. However, many parameters means many gates and thus large noise. One measure for a ``good'' quantum circuit is therefore to have as many parameters as necessary while being able to parametrize the entire state space of the simulated model. An optimal circuit taking this point of view would be minimal (there are no redundant parameters) and maximally expressive (the circuit can generate all relevant states). Parametric quantum circuits can be analyzed from this point of view using dimensional expressivity analysis (DEA)~\cite{Funcke2021} which we will review in \autoref{sec:DEA}. However, DEA can only tell us whether or not a given circuit is minimal and maximally expressive. The construction of a maximally expressive circuit is still highly non-trivial even though DEA does provide some guiding information. Additionally, having a maximally expressive circuit may not always be necessary or prudent for a given application, and DEA as described in~\cite{Funcke2021} does not provide any information on how useful a non-maximally expressive circuit is.

In this article, we will expand on the DEA by providing a method of constructing a minimal, maximally expressive circuit on $N+1$ qubits given a minimal, maximally expressive circuit on $N$ qubits in \autoref{sec:MMEC-construction}.

From \autoref{sec:best-approx} onwards we will move to non-maximally expressive circuits. These, by definition, cannot represent arbitrary states and, thus, the best case scenario is for a VQS to reproduce the best-approximation of the solution. In \autoref{sec:best-approx} we will define the necessary notation to discuss the worst-case error of the best-approximation. In particular, we will discuss Voronoi diagrams~\cite{Voronoi1908a,Voronoi1908b} to estimate the best-approximation error. Given a set of points $p_i$, a Voronoi diagram is a partition of space into regions. Each region is associated to one of the points $p_i$ and contains all points $x$ that are closer to $p_i$ than any other $p_j$. In other words, the Voronoi region associated with $p_i$ is the set of points whose best-approximation is~$p_i$. Knowing the Voronoi diagrams therefore allows us to compute the worst-case best-approximation error which is the largest distance a point $x$ can be from its corresponding best-approximation $p_i$, i.e., the largest possible distance between $x$ and $p_i$ where $x$ is an arbitrary point in the Voronoi region of $p_i$. This leads to an upper bound for the best-approximation error of a parametric quantum circuit by choosing the points $p_i$ to be a discrete set of sample states the given parametric quantum circuit can generate.

In \autoref{sec:full-Bloch} we will consider a minimal, maximally expressive circuit on the Bloch sphere as an example for the behavior of the here proposed best-approximation error estimates if applied to a maximally expressive circuit.  A first example of a non-maximally expressive circuit will be provided in \autoref{sec:first-example}. This example also highlights a potential problem if the best-approximation error is computed na\"ively. In \autoref{sec:practical-remarks} we will expand on this potential problem with some practical remarks. In particular, we will propose splitting the analysis into two steps. The first step will provide conditions that allow us to circumvent the obstacles identified in \autoref{sec:first-example} and provide a large lower bound on the best-approximation error should these conditions not be fulfilled. In the second step, once these conditions are fulfilled, the detailed analysis described in \autoref{sec:best-approx} is applicable without running into the potential problems observed in \autoref{sec:first-example}. In \autoref{sec:complexity} we will discuss the computational complexity of the best-approximation error estimate, in \autoref{sec:mapping} we propose an efficient mapping procedure that allows for exponential speed-up using a quantum device, and in \autoref{sec:scaling} we will discuss the scaling with respect to dimensionality. In \autoref{sec:image-volume} we will describe a relationship between the best-approximation error and the internal volume of the circuit image. This relationship will be illustrated with an example in \autoref{sec:spiral-bloch}, and we will discuss its impact on VQSs in \autoref{sec:local-optimizers}. Finally, we conclude the article in \autoref{sec:conclusion}.

\section{Dimensional Expressivity Analysis}\label{sec:DEA}
Within the context of dimensional expressivity analysis~\cite{Funcke2021}, we consider a parametric quantum circuit to be a continuously differentiable map $C:\ \Pp\to\Sp$. Here, $\Pp$ is the parameter space and assumed to be a (usually compact) manifold without boundary. For example, if only rotation gates \mbox{$R_G(\theta)=\exp\l(-i\frac\theta2 G\r)$} for gates $G$ with $G^2=1$ are used, then $\Pp$ is the flat torus $(\rn/2\pi\zn)^N$, where $N$ is the number of parameters in the circuit. $\Sp$ is the ``relevant'' state space, i.e., a submanifold of the unit sphere in the complex $2^Q$-dimensional Hilbert space for a $Q$-qubit quantum device.

The primary objective of the DEA is to identify redundant parameters in a given circuit $C$. Once these redundant parameters are identified, the parameter space $\Pp$ can be restricted by setting all redundant parameters to constants, and the thus restricted circuit is locally surjective to the image $\Mp$ of the original (unrestricted) circuit. This furthermore allows us to prove that $\Mp$ is locally a manifold and that the restricted circuit is minimal, i.e., removal of any further parameters (by setting them constant) will result in a restriction of the image~$\Mp$.

DEA identifies independent and redundant parameters inductively. The first parameter $\theta_1$ is always independent (unless the circuit is invariant under change of $\theta_1$ which is assumed to be not the case). Assuming that $\theta_1,\ldots,\theta_k$ are already identified as independent, we need to check $\theta_{k+1}$. The parameter $\theta_{k+1}$ is redundant if and only if the change in $C(\theta)$ given a perturbation of $\theta_{k+1}$ can also be achieved keeping $\theta_{k+1}$ fixed and varying $\theta_1,\ldots,\theta_k$ appropriately. In other words, $\theta_{k+1}$ is redundant if and only if the tangent vector~$\d_{\theta_{k+1}}C(\theta)$ is a linear combination of the tangent vectors $\d_{\theta_1}C(\theta),\ldots,\d_{\theta_k}C(\theta)$. Since our parameter space $\Pp$ is a real manifold, this linear combination is to be taken with respect to real coefficients. To test this linear independence, we consider the matrix
\begin{align}
  S_{k+1}=\l(\Re\langle\d_{\theta_m}C(\theta),\d_{\theta_n}C(\theta)\rangle\r)_{1\le m,n\le k+1}.
\end{align}
This matrix is invertible if and only if $\theta_{k+1}$ is independent. If $\theta_{k+1}$ is found to be redundant (i.e., $S_{k+1}$ is not invertible), then $\theta_{k+1}$ is removed as a parameter (i.e., it is kept constant) and the remaining parameters are tested after possibly re-labeling $\theta_j\mapsto \theta_{j-1}$ for $j>k+1$ in the redundant case.

Furthermore, the matrix $S_{k+1}$ can be measured efficiently using a quantum device. If all parametric gates are rotation gates $R_{G_n}(\theta_n)=\exp\l(-i\frac{\theta_n}{2} G_n\r)$ for some gate $G_n$ (which can be non-trivial like $\mathrm{CCNOT}$ or $ X_{q_1}X_{q_2}Y_{q_3}$, i.e., two $X$ gates acting on qubits $q_1$ and $q_2$ and a $Y$ gate acting on qubit~$q_3$ simultaneously), then $\d_{\theta_n}C(\theta)=-\frac{i}{2}\gamma_n$, where $\gamma_n$ is a quantum circuit with the additional gate $G_n$ inserted after the application of the gate $R_{G_n}(\theta_n)$. Hence, we obtain 
\begin{align}
  \Re\langle\d_{\theta_m}C(\theta),\d_{\theta_n}C(\theta)\rangle=\frac14\Re\langle\gamma_m,\gamma_n\rangle,
\end{align}
and $\Re\langle\gamma_m,\gamma_n\rangle$ can be measured on the quantum device.

Following the algorithm outlined in section 6 of~\cite{Funcke2021}, we add an ancilla qubit, which we initialize in $\frac{\ket0+\ket1}{\sqrt2}$. We then apply the gate sequence of our parametric quantum circuit but after application of $R_{G_m}$ and $R_{G_n}$ we insert the gates $X_{\mathrm{anc}}CG_mX_{\mathrm{anc}}$ and $CG_n$, respectively. Here, $CG_j$ denotes a controlled $G_j$ gate that uses the ancilla as control, and the pair of $X$ gates on the ancilla ($X_{\mathrm{anc}}$) ensures that $CG_m$ is conditioned on the ancilla being in $\ket0$. Finally, we apply another Hadamard gate to the ancilla. The state of the quantum device after all these steps is
\begin{align}
  \frac{\ket0\otimes(\ket{\gamma_m}+\ket{\gamma_n})+\ket1\otimes(\ket{\gamma_m}-\ket{\gamma_n})}{2},
\end{align}
and measuring the ancilla qubit yields the probability of finding the ancilla in $\ket0$
\begin{align}
  \mathrm{prob}(\mathrm{anc}=0) = \frac{1+\Re\langle\gamma_m,\gamma_n\rangle}{2}.
\end{align}

This algorithm of testing linear independence requires $\order((k+1)^2\epsilon^{-2})$ quantum device calls, where $\epsilon$ is the acceptable level of uncertainty for the measurement $\mathrm{prob}(\mathrm{anc}=0)$ (a~result of quantum device noise and finite statistics). The test for invertibility is classical and scales like $\order((k+1)^3)$, and the entire process needs to be repeated for $2\le k+1\le N$. Hence, this independence test algorithm scales like $\order(N^4\epsilon^{-2})$. On the quantum device it also requires an additional ancilla qubit and six additional gates.

It is important to note that this algorithm can be extended to compute any $\Re\langle C_1,C_2\rangle$ for circuits $C_1$ and $C_2$ by controlling all gates in $C_1$ and $C_2$ on an ancilla and ensuring that $C_1$ is conditioned on the ancilla being in $\ket0$ and $C_2$ is conditioned on the ancilla being in $\ket 1$. This will become important in \autoref{sec:complexity}, \autoref{sec:mapping}, and \autoref{sec:scaling}.

\section{Construction of minimal, maximally expressive circuits on $Q+1$ qubits}\label{sec:MMEC-construction}
DEA as described in \autoref{sec:DEA} requires a candidate circuit to be optimized. In many cases, finding such a candidate circuit is not an easy task. However, once a candidate is found, even if it is ``too expressive'' in the sense that it can generate states that are not physically meaningful, DEA provides a powerful tool. In particular, DEA can be used to remove symmetries such as invariance under a global phase transformation or even gauge invariance. In this section, we therefore want to discuss a means of generating candidate circuits. In order for this construction to be sufficiently general, we want the candidate circuits to be maximally expressive on $Q$ qubits. Furthermore, for the method to be efficient, we want the candidate circuits to be minimal in the number of parameters. Thus, if we can construct a minimal, maximally expressive circuit with an image manifold that still contains symmetries we wish to remove, then DEA can be used for the symmetry removal, while not having to remove any additional redundant parameters, which keeps the entire circuit construction process streamlined. 

Such streamlining can be achieved using a minimal, maximally expressive circuit for the entire $Q$-qubit quantum device state space. Of course, further streamlining can be achieved if some of the unwanted symmetries can already be removed at this stage, but in the most general case, in which all symmetries are to be removed via DEA, a minimal, maximally expressive circuit for the entire device state space is the optimal starting point. Since the $Q$-qubit state space is the unit sphere in the complex $2^Q$-dimensional Hilbert space, any minimal, maximally expressive circuit must have $2^{Q+1}-1$ real parameters. Starting with a single qubit, a minimal, maximally expressive circuit $C_1$ is given by
\begin{align}
  C_1(\theta)=R_Y(\theta_3)R_Z(\theta_2)R_X(\theta_1)\ket0.
\end{align}
The fact that this circuit is minimal and maximally expressive can be checked using DEA~\cite{Funcke2021}.

Given a $Q$-qubit minimal, maximally expressive circuit $C_Q$, we want to construct a minimal, maximally expressive circuit $C_{Q+1}$ on $Q+1$ qubits. The circuit $C_{Q+1}(\theta_1,\theta',\theta'')$ can be constructed by controlling the circuit $C_Q$ on the newly added qubit and considering
\begin{align*}
  \Qcircuit @C=1em @R=.7em {
    \lstick{\ket{0}} & \gate{R_X(\theta_1)} & \ctrl{1}    & \gate{X} & \ctrl{1}    & \qw\\
    \lstick{\ket{0}^{\otimes Q}} & \qw & \gate{C_Q(\theta')} & \qw & \gate{C_Q(\theta'')}&\qw\\
  }
\end{align*}
where $\theta'$ and $\theta''$ are a set of $2^{Q+1}-1$ parameters each. The new circuit $C_{Q+1}$ therefore has $2^{Q+2}-1$ parameters, i.e., it is minimal if and only if it is maximally expressive.

To show maximal expressivity of $C_{Q+1}$, we recall that any $(Q+1)$-qubit state $\ket\psi$ can be expressed as 
\begin{align}
  \ket\psi=\cos\frac{\theta_0}{2}\ket0\otimes \ket{\psi_0}+\sin\frac{\theta_0}{2}\ket1\otimes\ket{\psi_1}
\end{align}
with $Q$-qubit states $\ket{\psi_0}$ and $\ket{\psi_1}$. Through direct computation, we obtain
\begin{align}
  \begin{aligned}
    &C_{Q+1}(\theta_1,\theta',\theta'')\\
    =&\cos\frac{\theta_1}{2}\ket1\otimes C_Q(\theta'')-i\sin\frac{\theta_1}{2}\ket0\otimes C_Q(\theta'),
  \end{aligned}
\end{align}
and choosing $\theta_1=\pi-\theta_0$, $-iC_Q(\theta')=\ket{\psi_0}$, and $C_Q(\theta'')=\ket{\psi_1}$ shows that $C_{Q+1}$ is maximally expressive on $Q+1$ qubits if $C_Q$ is maximally expressive on $Q$ qubits (which we assumed by induction).

Finally, we need to consider the controlled circuit $CC_Q$. By induction, we assume that $C_Q$ contains only single-qubit gates and $\mathrm{CNOT}$ gates. Then, $CC_Q$ contains only controlled single-qubit gates and Toffoli gates. Controlled single-qubit gates can be constructed from $\mathrm{CNOT}$ and single-qubit gates, and the Toffoli gate can be constructed from $R_Y$ and $\mathrm{CNOT}$ gates~\cite{NielsenChuang}. Hence, $C_{Q+1}$ can be constructed using only $\mathrm{CNOT}$ and single-qubit gates.

It should also be noted that this construction can be extended to automatically remove some unwanted symmetries. For example, two states on the quantum device differing only by a phase factor are equivalent. Therefore, we may wish to remove global phase symmetry from the set of states the proposed circuit can generate. On a single qubit, such a circuit without global phase symmetry is given by \mbox{$C_1(\theta)=R_Z(\theta_2)R_X(\theta_1)\ket0$.} Similarly, this global phase symmetry can be incorporated into a circuit if we need it. To illustrate this, let us consider a $Q$-qubit circuit \mbox{$C_Q(\theta)=U(\theta)\ket\init$} and consider a gate set $U_\init$ such that $\ket\init=U_\init\ket0$. Then, we can incorporate global phase symmetry into $C_Q$ by adding an additional parameter $\phi$ and considering the circuit $\tilde C_Q(\phi,\theta)=U(\theta)U_\init R_Z(\phi) U_\init^*\ket\init$ where the $R_Z$ gate may act on any qubit. Choosing the \mbox{$Q$-qubit} initial state $\ket\init$ to be $\ket0$, we can adapt the construction described above by considering the $(Q+1)$-qubit circuit
\begin{align*}
  \Qcircuit @C=1em @R=.7em {
    \lstick{\ket{0}} & \gate{R_X(\theta_1)} & \ctrl{1}    & \gate{X} & \ctrl{1}    & \qw\\
    \lstick{\ket{0}^{\otimes Q}} & \gate{R_Z(\theta_2)} & \gate{C_Q(\theta')} & \qw & \gate{C_Q(\theta'')}&\qw\\
  }
\end{align*}
which generates the state 
\begin{align}
  \begin{aligned}
    C_{Q+1}(\theta_1,\theta_2,\theta',\theta'')=&\cos\frac{\theta_1}{2}\ket1\otimes C_Q(\theta'')\\
    &-i\sin\frac{\theta_1}{2}\ket0\otimes \tilde C_Q(\theta_2,\theta').
  \end{aligned}
\end{align}
This circuit $C_{Q+1}$ does not have a global phase symmetry as the choice of phase factor is fixed by $\cos\frac{\theta_1}{2}\ket1\otimes C_Q(\theta'')$.

\section{Best-approximation error}\label{sec:best-approx}
In \autoref{sec:MMEC-construction} we have shown how to generate minimal, maximally expressive circuits on arbitrarily many qubits. Should we wish to remove symmetries from the image $\Mp$, we can achieve this using DEA as well. This leaves us with the problem that sometimes maximally expressive circuits may not be feasible experimentally. In such circumstances, we want to use a circuit that is no longer maximally expressive and therefore may not be able to represent the solution of a given VQS. In this case, we want to estimate the worst-case error of a best-approximation for a given state using  a given non-maximally expressive circuit. 

Let us consider a parametric quantum circuit $C:\ \Pp\to\Sp$ mapping a parameter space $\Pp$ into a state space $\Sp$ that is no longer necessarily the entire unit sphere as we discussed in \autoref{sec:MMEC-construction}. Let $d:\ \Sp^2\to\rn_{\ge0}$ be a suitable distance in $\Sp$, e.g., the orthodromic (great-circle) distance as induced by the geometry of the quantum device state space, and define the distance of a point in $\Sp$ to $\Mp$ as
\begin{align}
  d_\Mp:\ \Sp\to\rn_{\ge0};\ x\mapsto \inf_{\theta\in\Pp}d(x,C(\theta)).
\end{align}
Thus, the quantum circuit $C$ is dense, i.e., it has a dense image, if and only if $\sup_{x\in\Sp}d_\Mp(x)=0$. We note that this is only possible if $\Pp$ is non-compact or $C$ is maximally expressive. As $\Pp$ is commonly a compact manifold, we note in such cases
\begin{align}
  d_\Mp(x) = 0\ \iff\ \ex \theta\in\Pp:\ x=C(\theta). 
\end{align}
In other words, if $\Pp$ is compact, then $C$ being dense is equivalent to $C$ being maximally expressive. Furthermore, if $\Pp$ is compact and $C$ is not maximally expressive, then 
\begin{align}
  \alpha_C:=\sup_{x\in\Sp}d_\Mp(x)
\end{align}
is strictly positive and denotes the largest distance that a point in $\Sp$ can be separated from~$\Mp$. $\alpha_C$ is therefore the best-approximation error of the circuit $C$ in the sense that 
\begin{itemize}
\item the best-approximation $C(\theta_x)$ of a given point $x\in\Sp$ satisfies $d(x,C(\theta_x))\le\alpha_C$ and
\item for all $\alpha$ with $0<\alpha<\alpha_C$ there exists an $x\in\Sp$ such that the best-approximation $C(\theta_x)$ of $x$ satisfies $d(x,C(\theta_x))>\alpha$. 
\end{itemize}
Hence, $\alpha_C$ is a measure of how good a given quantum circuit~$C$ approximates the state space $\Sp$.

In order to find an upper bound on $\alpha_C$, let us approximate $\Mp$ with a discrete point set $\Dp$. For example, we can choose $k\in\nn$ sufficiently large, uniformly sample values $\theta_1,\ldots,\theta_k\in\Pp$, and set $\Dp:=\{C(\theta_1),\ldots, C(\theta_k)\}$. We say that $\Dp$ is an $\eps$-dense approximation of $\Mp$ if and only if $\sup_{x\in\Mp}\inf_{y\in\Dp}d(x,y)<\eps$. Choosing an $\eps$-dense approximation~$\Dp$, we obtain
\begin{align}
  \alpha_C\le\sup_{x\in\Sp}\inf_{y\in\Dp}d(x,y)\le \alpha_C+\eps.
\end{align}
Our objective is therefore to compute $\sup_{x\in\Sp}\inf_{y\in\Dp}d(x,y)$ as an estimate for $\alpha_C$.

Let $\delta$ be a point in $\Dp$. The Voronoi region $R_\delta$ corresponding to $\delta$ is defined as the set
\begin{align}
  R_\delta:=\l\{x\in\Sp;\ d(x,\delta) = \min_{\delta'\in\Dp}d(x,\delta')\r\}.
\end{align}
Furthermore, let $V_\delta$ be the set of vertices of $R_\delta$. Then, 
\begin{align}
  \sup_{x\in\Sp}\inf_{y\in\Dp}d(x,y) = \max_{\delta\in\Dp}\max_{v\in V_\delta} d(\delta,v)
\end{align}
can be checked in finitely many steps and provides the estimate for $\alpha_C$ we were looking for. 

\section{A full Bloch sphere example}\label{sec:full-Bloch}
As an initial example, let us consider a circuit that is maximally expressive for the Bloch sphere. Given that the circuit in this section will be maximally expressive, this example should reproduce $\alpha_C=0$ and highlight the optimal behavior of the analysis.

\begin{figure}[!htp]
  \includegraphics[width=\columnwidth]{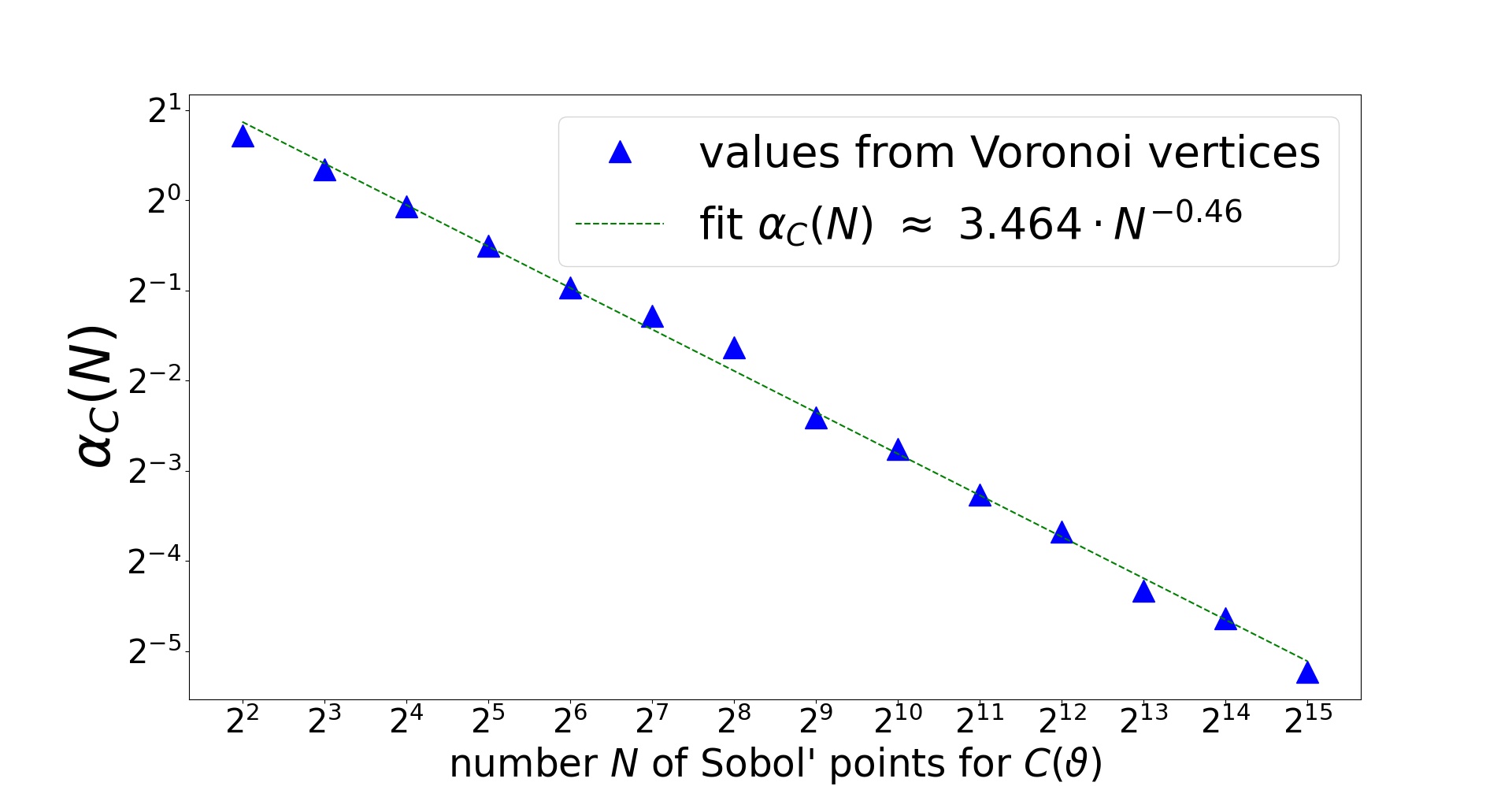}
  \caption{Estimate of $\alpha_C(N)$ for the circuit $C$  given in \autoref{eq:full_bloch_circuit} using $N$ scrambled Sobol' points.\\ The blue triangles show $\alpha_C(N)=\max_{\delta\in\Dp_N}\max_{v\in V_\delta} d(\delta,v)$ as obtained from spherical Voronoi diagrams. The dashed green line is a convergence rate fit of the form $\alpha_C(N)\propto N^{\rho}$. }\label{fig:alpha_C_Sobol}
\end{figure}

In this section, we will choose the representation of vectors~$\ket\psi$ on the Bloch sphere
\begin{align}\label{eq:bloch_rep}
  \ket\psi=\cos\frac{\Theta}{2}\ket0+e^{i\Phi}\sin\frac{\Theta}{2}\ket1
\end{align}
with $0\le\Theta\le\pi$ and $0\le\Phi<2\pi$, as well as, the circuit
\begin{align}\label{eq:full_bloch_circuit}
  \begin{aligned}
    C(\theta)=&R_Z(\theta_2)R_Y(\theta_1)\ket0\\
    \cong&\cos\frac{\theta_1}{2}\ket0+e^{i\theta_2}\sin\frac{\theta_1}{2}\ket1.
  \end{aligned}
\end{align}
From \autoref{eq:bloch_rep} and \autoref{eq:full_bloch_circuit}, we immediately observe that the circuit is maximally expressive -- in fact, a double cover for $0\le\theta_1\le2\pi$ -- and we may restrict $\theta_1$ to the interval $[0,\pi]$. The mapping of $\ket{\psi(\Theta,\Phi)}$ into $\rn^3$ is then given by
\begin{align}
  \ket{\psi(\Theta,\Phi)}\mapsto
  \begin{pmatrix}
    \sin\Theta\cos\Phi\\
    \sin\Theta\sin\Phi\\
    \cos\Theta
  \end{pmatrix}.
\end{align}
Using 
\begin{align}
  \alpha_C(N):=\max_{\delta\in\Dp_N}\max_{v\in V_\delta} d(\delta,v)
\end{align}
as an estimate for $\alpha_C$, where $\Dp_N=\{C(\theta_1),\ldots,C(\theta_N)\}$, and generating the $\theta_k$ using a scrambled Sobol' sequence in $[0,\pi]\times[0,2\pi]$, we expect to see $\alpha_C(N)\to0$ as $N\to\infty$. \autoref{fig:alpha_C_Sobol} shows that this is indeed the case and that the rate of convergence is comparable to Monte Carlo type convergence $\alpha_C(N)\approxpropto1/\sqrt{N}$. Furthermore, we will observe in \autoref{sec:scaling} that this rate of convergence (including the prefactor) is comparable to the theoretically optimal rate of convergence for the estimation of $\alpha_C$ based on Voronoi vertices.

\section{A first non-maximally expressive example}\label{sec:first-example}
In contrast to the maximally expressive example of \autoref{sec:full-Bloch}, we will now consider a non-maximally expressive circuit on the Bloch sphere. This circuit will represent a very bad choice of non-maximally expressive examples since it will only map to a great circle of the Bloch sphere. This has two important effects. Firstly, the worst-case best-approximation error will be very large; $\alpha_C=\frac\pi2$ in fact. Secondly, the computation of the Voronoi regions will encounter some non-trivial obstacle which we will discuss in more detail in \autoref{sec:practical-remarks}.

For ease of computation, we will consider the Bloch sphere embedded into $\rn^3$ using the real basis $\ket0$, $\ket1$, and $i\ket 1$. Furthermore, let us consider the circuit
\begin{align}
  C(\theta)=R_Y(\theta)\ket0=\cos\frac\theta2\ket0+\sin\frac\theta2\ket1. 
\end{align}
Then $\Mp$ is the great circle in the $\ket0$-$\ket1$-plane and consequently the worst-case best-approximation error is given by the distance of $\Mp$ to the poles $\pm i\ket1$, i.e., 
\begin{align}
  \alpha_C=d_\Mp(i\ket1)=\frac\pi2.
\end{align}
To compute $\alpha_C$ from a set $\Dp$ and its corresponding Voronoi vertices, we draw values 
\begin{align}
  0\le\theta_1<\theta_2<\ldots<\theta_k<2\pi
\end{align}
and consider the points 
\begin{align}
  C(\theta_k)=\cos\frac{\theta_k}{2}\ket0+\sin\frac{\theta_k}{2}\ket 1.
\end{align}
We assume that $k$ is large enough and that the $\theta_k$ are sufficiently well distributed such that 
\begin{align}
  \Dp=\{C(\theta_j);\ 1\le j\le k\}
\end{align}
is an $\eps$-dense approximation with $\eps<\pi$. For example, we may choose $k\ge3$ equidistant points $\theta_j=\frac{2\pi (j-1)}{k}$.

For $1<j<k$, the corresponding Voronoi region therefore is bounded by the great circles connecting $i\ket1$ and $-i\ket 1$, which pass through $\cos\frac{\theta_j+\theta_{j-1}}{4}\ket0+\sin\frac{\theta_j+\theta_{j-1}}{4}\ket 1$ and $\cos\frac{\theta_j+\theta_{j+1}}{4}\ket0+\sin\frac{\theta_j+\theta_{j+1}}{4}\ket 1$. Hence, we can set $V_{C(\theta_j)}$ to these four points and obtain 
\begin{align}
  \max_{v\in V_{C(\theta_j)}} d\l({C(\theta_j)},v\r) = d\l(C(\theta_j),i\ket1\r)=\frac{\pi}{2}.
\end{align}
For $V_{C(\theta_1)}$ and $V_{C(\theta_k)}$, we note that $C(2\pi)=-\ket0$ is equivalent to $C(0)=\ket0$. Thus, $V_{C(\theta_1)}$ and $V_{C(\theta_k)}$ contain the already discussed points $i\ket1$ and $-i\ket 1$, as well as, $\cos\frac{\theta_1+\theta_{2}}{4}\ket0+\sin\frac{\theta_1+\theta_{2}}{4}\ket 1$ and $\cos\frac{\theta_k+\theta_{k-1}}{4}\ket0+\sin\frac{\theta_k+\theta_{k-1}}{4}\ket 1$, respectively. Furthermore, both $V_{C(\theta_1)}$ and $V_{C(\theta_k)}$ contain the point between $C(\theta_1)$ and $C(\theta_k)$, i.e., 
\begin{align}
  C\l(\frac{2\pi+\theta_1+\theta_k}{2}\mod 2\pi\r).
\end{align}
Again, we conclude 
\begin{align}
  \max_{v\in V_{C(\theta_1)}} d\l({C(\theta_1)},v\r) = d\l(C(\theta_1),i\ket1\r)=\frac{\pi}{2}
\end{align}
and 
\begin{align}
  \max_{v\in V_{C(\theta_k)}} d\l({C(\theta_k)},v\r) = d\l(C(\theta_k),i\ket1\r)=\frac{\pi}{2}.
\end{align}
Hence, we have shown that
\begin{align}
  \alpha_C\le \max_{\delta\in\Dp}\max_{v\in V_\delta} d(\delta,v) = \frac\pi2\le \alpha_C+\eps
\end{align}
holds for all $\eps>0$ and thus
\begin{align}
  \alpha_C=\frac\pi2.
\end{align}

\section{Practical remarks}\label{sec:practical-remarks}
The computation of $\alpha_C$ relies on the computation of Voronoi diagrams. This may be tricky due to the nature of a lower dimensional circuit manifold $\Mp$. As the example above shows, this is relatively easy to see if the state space $\Sp$ is the entire unit sphere $\d B_{\cn^{2^Q}}$ of the $Q$-qubit Hilbert space $\cn^{2^Q}$ and if $\Mp$ is contained in a lower dimensional subspace $L$ of $\cn^{2^Q}$. The Voronoi regions then only have two proper vertices ($i\ket1$ and $-i\ket1$ in the case above) and we artificially added the mid-points between samples to properly define the Voronoi regions. This is likely to be difficult for an automated process to handle. However, if we use spherical Voronoi diagrams with samples such that the matrix $(C(\theta_1) \cdots C(\theta_k))$ has rank $2^{Q+1}$, then the Voronoi regions are well-defined on the sphere $\d B_{\cn^{2^Q}}$. Should $\Sp$ be a submanifold of the quantum device state space, then this may require further analysis of Voronoi diagrams in $\Sp$. However, if $\Mp$ is contained in a lower dimensional subspace~$L$ of the quantum device Hilbert space, then the poles normal to $L$ will be separated from $\Mp$ and therefore $\alpha_C\ge\frac\pi2$. This can be extended to state spaces $\Sp$ that are not the unit sphere by considering points in $\Sp$ that are not in $L$ and taking their distance as a lower bound for $\alpha_C$.

In many physical examples, we consider a state space $\Sp$ that is an intersection of a lower dimensional subspace $L$ of the quantum device Hilbert space and the unit sphere in the quantum device Hilbert space, e.g., if we are considering state spaces $\Sp$ that are eigenspaces of an operator describing a symmetry, such as the momentum $1$ sector for translational symmetry (cf., e.g., section 5 in \cite{Funcke2020}). Then,~$\Sp$ is the unit sphere in the lower dimensional subspace. In other words, all statements about spherical state spaces above hold. In particular, we need at least $\dim_\rn L$ points for the matrix $(C(\theta_1) \cdots C(\theta_k))$ to have rank $\dim_\rn L$. As long as $\rank(C(\theta_1) \cdots C(\theta_k))<\dim_\rn L$, we know that all points lie in a lower dimensional subspace intersected with~$\Sp$, which implies $\alpha_C\ge\frac\pi2$ considering the poles of $\Sp$ with respect to that lower dimensional space.

With respect to the problem of $\ket0\cong-\ket0$ we noticed in the example, we can circumvent this by artificially introducing a phase symmetry to the circuit. In other words, if \mbox{$C(\theta)=U(\theta)\ket{\init}$} and $\ket\init=U_\init\ket0$, then we may analyze 
\begin{align}
  \tilde C(\phi,\theta)=U(\theta)U_\init R_Z(\phi)U_\init^*\ket{\init}
\end{align}
instead.

Alternatively, for rotation gates \mbox{$R_G(\theta)=\exp\l(-i\frac{\theta}{2}G\r)$} with $G^2=1$, we could also consider $\theta\in[0,4\pi]$ instead of \mbox{$\theta\in[0,2\pi]$.}

\section{Complexity analysis}\label{sec:complexity}
The estimation of $\alpha_C$ on a spherical state space $\Sp$ has two primary components. First, we need to ensure that the sample matrix
\begin{align}
  \Sigma_N=
  \begin{pmatrix}
    \mid&\mid&&\mid\\
    \Re C(\theta_1)&\Re C(\theta_2)&\cdots&\Re C(\theta_N)\\
    \mid&\mid&&\mid\\
    \\
    \mid&\mid&&\mid\\
    \Im C(\theta_1)&\Im C(\theta_2)&\cdots&\Im C(\theta_N)\\
    \mid&\mid&&\mid\\
  \end{pmatrix}
\end{align}
has rank $\dim_\rn\Sp+1$. If this is not the case, then all $C(\theta_k)$ lie in the intersection of $\Sp$ with a linear space $L$ of dimension $\dim_\rn L\le\dim_\rn\Sp$. This implies that there exist points in~$\Sp$ that are orthogonal to $L$ and thus $\alpha_C(N)\ge\frac\pi2$. We can compute $\rank\Sigma_N$ by computing the rank of 
\begin{align}
  S_N=\Sigma_N^*\Sigma_N.
\end{align}
For this, we observe that the $(j,k)$-element of $S_N$ is given by
\begin{align}
  (S_N)_{j,k}=&\Re\langle C(\theta_j),C(\theta_k)\rangle
\end{align}
and note that the hybrid quantum-classical algorithm described in \autoref{sec:DEA} can be used to compute the $N\times N$-matrix $S_N=\Sigma_N^*\Sigma_N$ with $\order(N^2\epsilon^{-2})$ many QPU calls, where $\epsilon$ is the acceptable level of error on the QPU data. The rank of $S_N$ can then be computed classically, e.g., using QR decomposition in $\order(N^3)$. Overall, this step therefore costs $\order(N^3\epsilon^{-2})$.

Only once $\rank S_N=\dim_\rn\Sp+1$, we need to consider the Voronoi diagrams. To compute the Voronoi diagrams, we need to embed $\Sp$ into $\rn^{\dim_\rn\Sp+1}$ and compute the vectors $C(\theta_1),\ldots,C(\theta_N)$ using this \mbox{$\rn^{\dim_\rn\Sp+1}$-representation.} This step is discussed in \autoref{sec:mapping}. Once this embedding is achieved, computing the Voronoi diagrams is efficient with worst-case cost in $\order\l(N^{\l\lceil\frac{\dim_\rn\Sp}{2}\r\rceil }\r)$ via stereographic projection and Delaunay triangulation~\cite{Fortune2017}. Since the Voronoi vertices are the circumcenters of Delaunay simplices, the number of Voronoi vertices coincides with the number of Delaunay simplices of $N$~points in $D=\dim_\rn\Sp+1$ dimensions, i.e., $\order(N^{\lceil D/2\rceil})$. Computing $\alpha_C(N)$ inefficiently via
\begin{align}
  \alpha_C(N)=\max_{v\in V}\min_{1\le j\le N}d(v,C(\theta_j)),
\end{align}
where $V$ is the entire set of Voronoi vertices -- i.e., by comparing all Voronoi vertices to all sample points rather than restricting the comparison only to vertices of each sample's Voronoi region -- we obtain a worst-case cost in $\order(N^{\lceil D/2\rceil+1})$ to compute $\alpha_C(N)$.

\section{Mapping $C(\theta)$ efficiently into $\rn^{\dim_{\rn}\Sp+1}$}\label{sec:mapping}
In order to efficiently map $C(\theta)$ into into $\rn^{\dim_{\rn}\Sp+1}$, we need to employ the quantum device again. But first, since we are working with real vector spaces, we cannot use $C(\theta)$ directly. Instead, we need to consider the real vector 
\begin{align}
  C_\rn(\theta):=
  \begin{pmatrix}
    \Re C(\theta)\\\Im C(\theta)
  \end{pmatrix}.
\end{align}
Hence, we are looking to map the sequence of samples $C_\rn(\theta_j)\in\rn^{2^{Q+1}}$ efficiently into $\rn^{\dim_{\rn}\Sp+1}$. To make this mapping efficient, we will map the set $\{C_\rn(\theta_j);\ j\le k\}$ into the subspace $\lin\{e_j;\ j\le r\}$, where $\{e_j;\ j\le \dim_{\rn}\Sp+1\}$ denotes the canonical basis of $\rn^{\dim_{\rn}\Sp+1}$ and 
\begin{align}
  r=\rank
  \begin{pmatrix}
    C_\rn(\theta_1)&C_\rn(\theta_2)&\ldots&C_\rn(\theta_k)
  \end{pmatrix}.
\end{align}
$C_\rn(\theta_1)$ is automatically mapped to 
\begin{align}
  v_1=
  \begin{pmatrix}
    1\\0\\0\\\vdots\\0
  \end{pmatrix}.
\end{align}
Then, we map $C_\rn(\theta_2)$ to 
\begin{align}
  v_2=
  \begin{pmatrix}
    v_{2,1}\\v_{2,2}\\0\\\vdots\\0
  \end{pmatrix}
\end{align}
with
\begin{align}
  v_{2,1}=&\langle v_2,v_1\rangle=\Re\langle C(\theta_2),C(\theta_1)\rangle,
\end{align}
which we can compute efficiently on the quantum device, and 
\begin{align}
  v_{2,2}=\sqrt{1-v_{2,1}^2}.
\end{align}
At this point, we need to keep track of the current subspace basis $B_k$ after mapping $C(\theta_1),\ldots, C(\theta_k)$. The initial basis~$B_1$ contains only the vector $v_1$. The vector $v_2$ is an element of $B_k$ with $k\ge2$ if and only if $v_{2,2}\ne0$.

If $v_2\notin B_2$, then $B_2=\{v_1\}$ and $C(\theta_3)$ is mapped to
\begin{align}
  v_3=
  \begin{pmatrix}
    v_{3,1}\\v_{3,2}\\0\\\vdots\\0
  \end{pmatrix}
\end{align}
with
\begin{align}
  v_{3,1}=&\Re\langle C(\theta_3),C(\theta_1)\rangle
\end{align}
and 
\begin{align}
  v_{3,2}=\sqrt{1-v_{3,1}^2}.
\end{align}
Similarly, $v_3$ is added to $B_3$ if and only if $v_{3,2}\ne0$.

On the other hand, if $v_2\in B_2$, then $C(\theta_3)$ is mapped to
\begin{align}
  v_3=
  \begin{pmatrix}
    v_{3,1}\\v_{3,2}\\v_{3,3}\\0\\\vdots\\0
  \end{pmatrix}
\end{align}
with
\begin{align}
  \begin{aligned}
    v_{3,1}=&\langle v_3,v_1\rangle=\Re\langle C(\theta_3),C(\theta_1)\rangle,\\
    v_{3,2}=&\frac{\Re\langle C(\theta_3),C(\theta_2)\rangle-v_{3,1}v_{2,1}}{v_{2,2}},\\
    v_{33}=&\sqrt{1-v_{3,1}^2-v_{3,2}^2},
  \end{aligned}
\end{align}
and $v_3$ is added to $B_3$ if and only if $v_{3,3}\ne0$.

In order to map $C(\theta_k)$, we have already mapped $C(\theta_1),\ldots,C(\theta_{k-1})$ and identified a basis $B_{k-1}$ with $b_{k-1}$ elements. We denote the elements of $B_{k-1}$ by $v_{\beta_1},\ldots,v_{\beta_{b_{k-1}}}$. Then $C(\theta_k)$ is mapped to
\begin{align}
  v_k=
  \begin{pmatrix}
    v_{k,1}\\\vdots\\v_{k,b_{k-1}+1}\\0\\\vdots\\0
  \end{pmatrix}
\end{align}
with
\begin{align}
  \begin{aligned}
    v_{k,1}=&\Re\langle C(\theta_k),C(\theta_{\beta_1})\rangle,\\
    v_{k,2}=&\frac{\Re\langle C(\theta_k),C(\theta_{\beta_2})\rangle-v_{k,1}v_{{\beta_2},1}}{v_{{\beta_2},2}},\\
    \vdots&\\
    v_{k,b_{k-1}}=&\frac{\Re\langle C(\theta_k),C(\theta_{\beta_{b_{k-1}}})\rangle-\sum_{j=1}^{b_{k-1}-1}v_{k,j}v_{\beta_{b_{k-1}},j}}{v_{\beta_{b_{k-1}},b_{k-1}}},\\
    v_{k,b_{k-1}+1}=&\sqrt{1-\sum_{j=1}^{b_{k-1}}v_{k,j}^2}.
  \end{aligned}
\end{align}
This is possible since by construction $v_{\beta_j,j}\ne0$. Of course, $v_k$ is added to $B_k$ if and only if $v_{k,b_{k-1}+1}\ne0$.

Combining this mapping of quantum states with the complexity analysis of \autoref{sec:complexity} shows that $\alpha_C$ can be estimated efficiently.

\section{Scaling for spherical state spaces and maximally expressive circuits}\label{sec:scaling}
Now that we know that the Voronoi estimation of $\alpha_C$ is efficiently implementable, the question becomes how many points are necessary for the Voronoi estimation to yield a sufficiently reliable estimate of $\alpha_C$.

The optimal scaling for $\alpha_C(N)$ on a $(D-1)$-dimensional sphere would be given by equidistant points. As such a configuration is not possible for arbitrary values of $N$, we must approximate this type of scaling. Considering spherical coordinates in $D-1$ dimensions
\begin{align}
  \begin{aligned}
    x_1=&\cos\phi_1\\
    x_2=&\sin\phi_1\cos\phi_2\\
    x_3=&\sin\phi_1\sin\phi_2\cos\phi_3\\
    \vdots&\\
    x_{D-1}=&\sin\phi_1\sin\phi_2\cdots\sin\phi_{D-2}\cos\phi_{D-1}\\
    x_{D}=&\sin\phi_1\sin\phi_2\cdots\sin\phi_{D-2}\sin\phi_{D-1},
  \end{aligned}
\end{align}
we choose circles of constant $\phi_1$ which are spaced a distance~$d_1$ apart on the surface of the sphere. Then, we discretize $\phi_2$ on each of the $\phi_1$ circles such that the $\phi_2$ values are spaced a distance $d_2$ apart ($d_2$ depends on the specific circle of constant $\phi_1$) and each value of $d_2$ is as close as possible to $d_1$. Iteratively, we can space $\phi_k$ for each chosen $\phi_1,\ldots,\phi_{k-1}$ such that the corresponding distances satisfy $d_1\approx d_2\approx\ldots\approx d_k$ as closely as possibly. As such, the volume of the sphere attributed to each point is approximately $\prod_{j=1}^{D-1}d_j$, i.e., approximately $d_1^{D-1}$. Since the surface area of a $(D-1)$-dimensional unit sphere is $\frac{2\pi^{D/2}}{\Gamma(D/2)}$, this means we can place $N\approx\frac{2\pi^{D/2}}{\Gamma(D/2)d_1^{D-1}}$ points. Conversely, although we wish to place $N$ points, we may not be perfectly able to, but we can place $N'\approx N$ points by choosing $d_1$ to be a fraction of $\pi$ as close as possible to $\l(\frac{2\pi^{D/2}}{\Gamma(D/2)N}\r)^{1/(D-1)}$. Each sample can then be interpreted as the approximate center of a cube with side length $\l(\frac{2\pi^{D/2}}{\Gamma(D/2)N}\r)^{1/(D-1)}$, which means the maximal distance that an arbitrary point can be away from any of these sample points is approximately $\l(\frac{2\pi^{D/2}}{\Gamma(D/2)N}\r)^{1/(D-1)}\frac{\sqrt{D}}{2}$. In other words, the best case scaling we can expect for $\alpha_C(N)$ with $D=\dim_\rn\Sp+1$ is
\begin{align}
  \alpha^\opt_C(N)=\l(\frac{2\pi^{D/2}}{\Gamma(D/2)N}\r)^{1/(D-1)}\frac{\sqrt{D}}{2}.
\end{align}
Returning to the Bloch sphere example, we have $D=3$, i.e., 
\begin{align}
  \alpha^\opt_C(N)=&\sqrt{3\pi}N^{-\frac12} \approx3.07 N^{-\frac12}.
\end{align}
A comparison with the observed $\alpha_C(N)\approx3.46N^{-.46}$ using a Sobol' sequence to generate samples $C(\theta_k)$ in \autoref{fig:alpha_C_Sobol} shows that the Sobol' point approximation is almost optimal. 

\section{$\alpha_C$ and the volume of lower dimensional $\Mp$}\label{sec:image-volume}
The scaling analysis of \autoref{sec:scaling} is optimized for spherical state spaces and maximally expressive circuits. If we wish to use a circuit $C$ with lower dimensional image $\Mp$ and guarantee a chosen best-approximation error $\alpha_C$, then we need to compare $\alpha_C$ to $\vol(\Mp)$ and observe that $\Mp$ has to have a large volume. In this section, we want to approximate a lower bound for $\vol(\Mp)$ in terms of $\alpha_C$.

First, we note that $\dim\Mp=\dim\Sp-1$ is the best case scenario (in terms of smallest possible volume of $\Mp$) since the projection onto a $\dim\Mp+1$ dimensional subspace of $\Sp$ reduces the distance between points and thus, reduces the volume of $\Mp$ as well as $\alpha_C$. After a suitable choice of coordinate transformation, we may therefore assume that $C(\theta)$ takes the form
\begin{align}
  C(\theta)=
  \begin{pmatrix}
    \cos(\gamma_1(\theta_1))\\
    \sin(\gamma_1(\theta_1))\cos(\gamma_2(\theta_1))\\
    \sin(\gamma_1(\theta_1))\sin(\gamma_2(\theta_1))\sigma(\theta_2,\ldots,\theta_n)
  \end{pmatrix},
\end{align}
where $\sigma$ is surjective onto the $(\dim\Mp-1)$-dimensional sphere, and $(\gamma_1,\gamma_2)$ is the path $C$ takes in the projection of $\Sp$ onto the $2$-dimensional subspace of $\Sp$ orthogonal to the sphere $\sigma$ maps to. 

To estimate the minimal volume of $\Mp$, we can use a greedy algorithm to construct a path $\gamma$ on the $2$-dimensional sphere that is $\alpha_C$ close to all points. For example, we may start at the north pole and move south for a distance $2\alpha_C$. Then, we move east along the reached latitude stopping $2\alpha_C$ short of a full revolution. Now we move south again for $2\alpha_C$ and then move east stopping $2\alpha_C$ short of a full revolution. We continue this process until we reach the south pole. For simplicity, we assume that $2\alpha_C$ is an integer fraction of $\pi$, i.e., $2\alpha_C=\frac{\pi}{n}$. Then, this path contains $n$ sections of going south for $2\alpha_C$ and $n-1$ sections of moving east for a full revolution but stopping short $2\alpha_C$. The length of this path $\ell_1$ is therefore
\begin{align}
  \ell_1=&2\alpha_C+\sum_{j=1}^{n-1}2\pi\sin\frac{\pi j}{n}=\frac{\pi}{n}+2\pi\cot\frac{\pi}{2n}.
\end{align}
Finally, we obtain the corresponding volume
\begin{align}
  V_1=&\ell_1\vol S^{\dim\Mp-1}
  =\frac{4\pi^{\frac{\dim\Mp}{2}}\l(\alpha_C+\pi\cot\alpha_C\r)}{\Gamma\l(\frac{\dim\Mp}{2}\r)}.
\end{align}

Another possible greedy algorithm is to use a spiral with $\gamma_1(\theta_1)=\pi\theta_1$ and $\gamma_2(\theta_1)=2\pi n\theta_1$ where $\theta_1\in[0,1]$. This spiral wraps $n$ times around the sphere and therefore has $2\alpha_C\approx \frac{\pi}{n}$ for large $n$. We may compute the length of this curve as
\begin{align}
  \ell_2=&\int_0^1\norm{\gamma'(t)}dt
  =\sqrt{\pi^2+4\pi^2n^2}
\end{align}
and thus the corresponding volume
\begin{align}
  V_2=&\ell_2\vol S^{\dim\Mp-1}\approx\frac{2\pi^{\frac{\dim\Mp}{2}+1}\sqrt{1+\frac{\pi^2}{\alpha_C^2}}}{\Gamma\l(\frac{\dim\Mp}{2}\r)}.
\end{align}
For $\alpha_C\ll1$, we can use $\cot(\alpha_C)=\frac{1}{\alpha_C}+\order(\alpha_C)$ to observe 
\begin{align}
  \frac{V_1}{V_2}\approx&\frac{2\cot\alpha_C}{\sqrt{1+\frac{\pi^2}{\alpha_C^2}}}\approx\frac{2}{\pi},
\end{align}
i.e., both estimates are of the same order of magnitude, and given the volume of the circuit image~$\Mp$, we can re-arrange the $V_1$ estimate to obtain
\begin{align}
  \alpha_C\gtrsim\frac{4\pi^{\frac{\dim\Mp}{2}+1}}{\Gamma\l(\frac{\dim\Mp}{2}\r)\vol(\Mp)}.
\end{align}

Of course, for this relation to be useful, we either need to compute $\alpha_C$, e.g., from Voronoi diagrams, or $\vol(\Mp)$. If we wish to compute $\vol(\Mp)$, then we may choose a quadrature rule with points $\theta_j$ and weights $\omega_j$ and estimate
\begin{align}
  \begin{aligned}
    \vol(\Mp)=&\int_{\Pp}\sqrt{\det g(\theta)}d\vol_{\Pp}(\theta)\\
    \approx&\sum_j\omega_j\sqrt{\det g(\theta_j)},
  \end{aligned}
\end{align}
where 
\begin{align}
  g_{jk}(\theta)=\Re\langle \d_jC(\theta),\d_kC(\theta)\rangle,
\end{align}
for which we already have an efficient algorithm in \autoref{sec:DEA}. It is important to note that $g$ is the same matrix as $S_N$ in \autoref{sec:DEA}. Thus, for $\det g$ to not be zero, the vectors $\d_j C(\theta)$ need to be linearly independent. In other words, the volume element $\sqrt{\det g(\theta)}d\vol_{\Pp}(\theta)$ needs to be computed with a minimal circuit.

\section{A Bloch sphere spiral example}\label{sec:spiral-bloch}
To illustrate this relationship between $\alpha_C$ and $\vol(\Mp)$, let us consider the family of circuits
\begin{align}\label{eq:bloch_spiral}
  \begin{aligned}
    C_n(\theta)=&R_Z(n\theta)R_Y(\theta)\ket0\\
    =&\cos\frac{\theta\mod\pi}{2}\ket0+e^{2in\theta}\sin\frac{\theta\mod\pi}{2}\ket1
  \end{aligned}
\end{align}
on the Bloch sphere with $\theta\in[0,2\pi]$. Here the ``$\mathrm{mod}\ \pi$'' condition simply ensures that the coefficient in front of $\ket0$ is non-negative as per our representation of the Bloch sphere introduced in \autoref{sec:full-Bloch}. In terms of the Bloch sphere representation 
\begin{align}
  \cos\frac{\Theta}{2}\ket0+e^{i\Phi}\sin\frac{\Theta}{2}\ket1\mapsto
  \begin{pmatrix}
    \sin\Theta\cos\Phi\\
    \sin\Theta\sin\Phi\\
    \cos\Theta
  \end{pmatrix},
\end{align}
$C_n(\theta)$ is represented by the vector
\begin{align}
  \gamma(\theta)=
  \begin{pmatrix}
    \sin(\theta\mod\pi)\cos(2n\theta)\\
    \sin(\theta\mod\pi)\sin(2n\theta)\\
    \cos(\theta\mod\pi)
  \end{pmatrix},
\end{align}
and we obtain
\begin{align}
  \det g(\theta)=&1+4n^2\sin(\theta\mod\pi)^2,
\end{align}
i.e.,
\begin{align}
  \vol(\Mp_n)=&\int_0^{2\pi}\sqrt{\det g(\theta)}d\theta=4E(-4n^2)
\end{align}
using the elliptic integral 
\begin{align}
  E(m)=\int_0^{\frac\pi2}\sqrt{1-m\sin(\theta)^2}d\theta.
\end{align}

\begin{figure}[!htp]
  \includegraphics[width=\columnwidth]{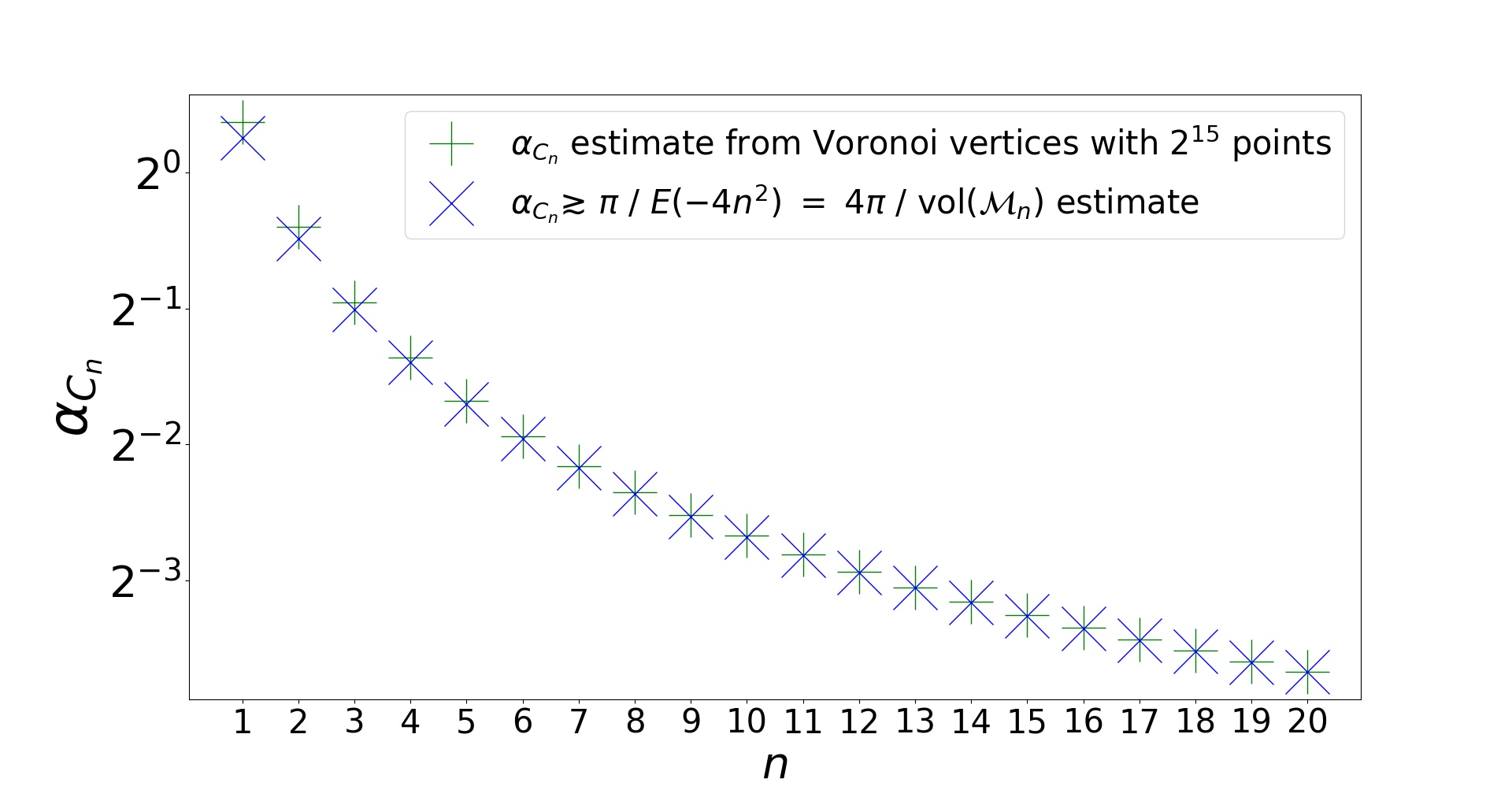}
  \caption{Estimate of $\alpha_{C_n}$ for the circuit $C$ given in \autoref{eq:bloch_spiral} using the volume estimate and the Voronoi diagram estimate. The Voronoi diagrams use $2^{15}$ samples generated from scrambled Sobol' points. }\label{fig:spiral-bloch}
\end{figure}

As in this case $\dim(\Mp_n)=1$, the estimated lower bound for $\alpha_{C_n}$ becomes
\begin{align}
  \alpha_{C_n}\gtrsim&\frac{4\pi^{\frac{3}{2}}}{\Gamma\l(\frac{1}{2}\r)\vol(\Mp_n)}=\frac{4\pi}{\vol(\Mp_n)}=\frac{\pi}{E(-4n^2)}.
\end{align}
Since $C_n$ is a ``spiral'' circuit that is also taking into account that the parameter space $\Pp$ should be a manifold without boundary (here $\rn/2\pi\zn$), we expect to see strong agreement between $\alpha_{C_n}$ estimated using $\frac{\pi}{E(-4n^2)}$ and $\alpha_{C_n}$ extracted from Voronoi diagrams. \autoref{fig:spiral-bloch} confirms this expectation. 

\section{Voronoi analysis for initializations of VQS with local optimizers}\label{sec:local-optimizers}
The example in \autoref{sec:spiral-bloch} and the estimate 
\begin{align}
  \alpha_C\vol(\Mp)\gtrsim\frac{4\pi^{\frac{\dim\Mp}{2}+1}}{\Gamma\l(\frac{\dim\Mp}{2}\r)}
\end{align}
show that quantum circuits that are not maximally expressive can be able to approximate points in the state space $\Sp$ to a given precision $\alpha_C$ at the cost of having large $\vol(\Mp)$. In particular, the spiral example shows that points that are close in state space $\Sp$ may have best-approximations with parameter values $\theta$ that are far apart in parameter space $\Pp$. This can be illustrated using the circuit 
\begin{align}
  \begin{aligned}
    C_n(\theta)=&R_Z(n\theta)R_Y(\theta)\ket0\\
    =&\cos\frac{\theta\mod\pi}{2}\ket0+e^{2in\theta}\sin\frac{\theta\mod\pi}{2}\ket1
  \end{aligned}
\end{align}
and a point $p$ half-way between points $C_n(\theta)$ and $C_n(\theta')$ with $\theta'=\theta+\frac{2\pi}{n}$. Any arbitrarily small perturbation of $p$ towards $C_n(\theta)$ will have best-approximation parameters close to $\theta$, whereas a perturbation towards $C_n(\theta')$ has best-approximation parameters close to $\theta'$, i.e., an entire revolution around the sphere later. If this point $p$ is equatorial, then arbitrarily small perturbations in $p$ yield best-approximations that are a distance approximately $2\pi$ apart as measured in $\Mp$. Considering an energy-type cost function
\begin{align}
  \Cost(\theta)=\bra{ C_n(\theta)}H\ket{C_n(\theta)}
\end{align}
for some Hamiltonian $H$ and supposing $n$ large, $\Mp_n$ is packed tightly around the solution state $\ket\psi$ that needs to be approximated by $C_n(\theta)$. However, this implies $\Cost(\theta)$ has many local minima corresponding to states $C_n(\theta)$ that are close to $\ket\psi$ in $\Sp$, but these minima are far away from each other as measured in the metric of $\Mp$. Hence, having an initial guess $C_n(\theta)$ close to $\ket\psi$ in $\Sp$ is not sufficient for a local optimizer to guarantee convergence to the best-approximation $C_n(\theta_\psi)$ of $\ket\psi$. Instead, the initial guess $C_n(\theta)$ must satisfy $\theta\approx\theta_\psi$ in parameter space as well.

The Voronoi analysis provides precisely such a set of initial guesses. Since every point of the state space $\Sp$ is $\alpha_{C_n}$ close to one of the Voronoi sample points, we can use the Voronoi sample points as initial guesses and run a local optimizer VQS for each of them (or at least those that have near optimal cost function values). The best-approximation will then be selected as the $C(\theta_\psi)$ among the solution candidates $C(\theta_j)$ provided by each VQS for which $\Cost(C(\theta_\psi))=\min_j\Cost(C(\theta_j))$.

\section{Conclusion}\label{sec:conclusion}
In this article we have extended previous work on dimensional expressivity analysis (DEA)~\cite{Funcke2021} in two important directions. First, we reviewed DEA and highlighted that it requires to have a candidate circuit to be optimized. However, finding a candidate circuit is not always straightforward. Hence, in \autoref{sec:MMEC-construction} we proposed an inductive process of constructing minimal, maximally expressive circuits on arbitrarily many qubits. This can be used directly, if the entire quantum device state space needs to be parameterized, or it may be optimized using DEA if certain symmetries are to be removed.

The second major open question surrounding DEA is related to non-maximally expressive circuits. Previously, DEA made no statements about non-maximally expressive circuits. Hence, we considered the worst-case best-approximation error and proposed estimating it using Voronoi vertices in \autoref{sec:best-approx}. This was supplemented with examples highlighting the optimal behavior of the analysis in \autoref{sec:full-Bloch} and some possible obstacles in \autoref{sec:first-example}, a discussion of practical aspects surrounding these obstacles in \autoref{sec:practical-remarks}, and an example applying the Voronoi estimation of the best-approximation error on a maximally expressive circuit in \autoref{sec:practical-remarks}.

In \autoref{sec:complexity} and \autoref{sec:mapping} we then discussed the computational complexity of the proposed Voronoi diagram 
based estimation of the best-approximation error using a hybrid quantum-classical algorithm that allows us to map the quantum states efficiently into classical memory and compute the Voronoi vertices. In particular, the quantum part of the algorithm scales quadratically in both the number of sample points~$N$ required for the Voronoi estimates and the inverse acceptable level of noise $\epsilon^{-1}$ coming from the quantum device. The algorithm without the Voronoi vertices computation is still in $\order(N^3\epsilon^{-2})$. The bottlenecks of the algorithm are therefore the computation of the Voronoi vertices which, in $D=\dim_\rn\Sp+1$ dimensions, is worst case in $\order(N^{\lceil D/2\rceil})$, and testing all $\order(N^{\lceil D/2\rceil})$ Voronoi vertices against (worst case) all Voronoi sample points is in $\order(N^{\lceil D/2\rceil+1})$.

In \autoref{sec:scaling} we then discussed the necessary number of points $N$ required to estimate the best-approximation error to a chosen level of accuracy. The optimal scaling  is proportional to $N^{\frac{1}{D-1}}$ and achieved for maximally expressive circuits. For non-maximally expressive circuits, as discussed in \autoref{sec:image-volume}, the internal volume of the set of reachable states $\Mp$ by the quantum circuit needs to be considered. In the same section, an efficient hybrid quantum-classical algorithm for the estimation of $\vol(\Mp)$ is also provided. Furthermore, we tested the relationship between the best-approximation error and $\vol(\Mp)$ using an example of a spiral circuit image $\Mp$ on the Bloch sphere in \autoref{sec:spiral-bloch}. Finally, we have discussed some of the expected obstacles when applying non-maximally expressive circuits to variational quantum simulations with local optimizers in \autoref{sec:local-optimizers}.

The work presented here therefore extends DEA~\cite{Funcke2021} by multiple practical aspects. We have provided minimal, maximally expressive candidate circuits that can be optimized using DEA on arbitrarily many qubits, which can further be optimized by removing unwanted symmetries via DEA. Should we choose to employ non-maximally expressive circuits, then we have provided the means of estimating the worst-case best-approximation error a priori to any VQS. In particular, if we want to ensure a certain best-approximation error $\alpha_C$, then we can predict the minimal number of samples necessary to construct a set of VQS initial states, such that every element of the relevant state space is $\alpha_C$ close to at least one point in the VQS initialization set. The VQS can then be used to optimize any of these initial states to find the best-approximation of the VQS solution for non-maximally expressive circuits.

\end{document}